\documentclass[12pt]{article}
\usepackage[cp1251]{inputenc}
\usepackage[english]{babel}
\usepackage{mathtext,amsfonts,euscript,graphicx,amsmath,amssymb,color}
\usepackage{cite}
\begin{document}
\selectlanguage{english}

\title{Steady-state composition of a two-component gas bubble growing in a liquid solution: self-similar approach}
 
\author{G. Yu. Gor\footnote{Current address: \textit{Department of Chemical and Biochemical Engineering, Rutgers, The State University of New Jersey, 98 Brett road, Piscataway, New Jersey 08854-8058, USA.} e-mail: \textit{gennady\_gor@mail.ru}} ~and A. E. Kuchma}

\maketitle

\begin{center}
\textit{Department of Theoretical Physics, Research Institute of Physics,\\
Saint-Petersburg State University,\\
1 Ulyanovskaya Street, Petrodvorets, St. Petersburg, 198504, Russia}
\end{center}

\begin{abstract}
\noindent
The paper presents an analytical description of the growth of a two-component bubble in a binary liquid-gas solution. We obtain asymptotic self-similar time dependence of the bubble radius and analytical expressions for the non-steady profiles of dissolved gases around the bubble. We show that the necessary condition for the self-similar regime of bubble growth is the constant, steady-state composition of the bubble. The equation for the steady-state composition is obtained. We reveal the dependence of the steady-state composition on the solubility laws of the bubble components. Besides, the universal, independent from the solubility laws, expressions for the steady-state composition are obtained for the case of strong supersaturations, which are typical for the homogeneous nucleation of a bubble.
\end{abstract}

\section{Introduction}
\label{sec:Introduction}

The subject of this paper is theoretical treatment of growth via diffusion of a two-component gas bubble in a supersaturated gas-liquid solution. Understanding the regularities of gas bubble growth in gas-liquid solutions is crucial for the study of various natural phenomena and for the management of different technological processes. Gas bubble growth is important e. g. for glass refining processes \cite{Cable-Frade}, polymeric foams production \cite{Kim-Kang-Kwak}. The growth of gas bubbles dissolved in magma is a process which governs volcanic eruptions \cite{Sparks-1978, Sahagian-1999, Lensky-Navon-Lyakhovsky-2004, L'Heureux-2007}.

The problem of diffusional bubble growth with single gas in it was first described theoretically in the classical paper by Epstein and Plesset \cite{Epstein-Plesset} and remains topical till nowadays when treated under different specific conditions \cite{Lastochkin-Favelukis, Divinis-2004, Divinis-2005, Kostoglou-Karapantsios-2007, Grinin-Kuni-Gor, Kuchma-Gor-Kuni, Gor-Kuchma-Sievert}.

While the discussion in Refs. \cite{Cable-Frade, Kim-Kang-Kwak, Lensky-Navon-Lyakhovsky-2004, L'Heureux-2007, Epstein-Plesset, Lastochkin-Favelukis, Divinis-2004, Divinis-2005, Kostoglou-Karapantsios-2007, Grinin-Kuni-Gor, Kuchma-Gor-Kuni, Gor-Kuchma-Sievert}
 is limited to bubbles containing only one gas, often there are two or more different gases dissolved in the solution and, therefore, presenting in the bubble. The growth of a multicomponent bubble in glass melts was described theoretically by Ramos \cite{Ramos-Multi} and Cable and Frade \cite{Cable-Frade-Multi}. The main question in these studies was the time dependence of multicomponent bubble radius. In Ref. \cite{Ramos-Multi}
  the growth was considered from a small initial size (when Laplace pressure in the bubble is of the order of the external pressure of the solution or more, and therefore, the surface tension strongly influences bubble growth); and there the problem was treated numerically only. However, Ramos \cite{Ramos-Multi} reported a small effect of surface tension on bubble growth.

Cable and Frade \cite{Cable-Frade-Multi} neglected the effect of the surface tension initially. They solved the problem numerically for the arbitrary (but large enough to neglect Laplace pressure) initial radius and also presented an analytical description of the growth from the zero initial radius. Ref. \cite{Cable-Frade-Multi}
 states that the composition of a multicomponent bubble tends to a steady-state value; however, explicit equations for calculating this composition have not been obtained there. (In Ref. \cite{Cable-Frade-Multi}
  this value of the gas composition in the bubble was erroneously called "equilibrium". If the composition was equilibrium, the bubble would be unable to grow). Besides that, in Refs. \cite{Ramos-Multi, Cable-Frade-Multi}
   the discussion was limited to low supersaturations typical for problems related with glass production. Finally, the influence of gas solubility laws on the steady composition of the bubble was totally excluded from discussion.

In the current paper we will consider a bubble which consists of two different gases in a liquid solution where both of these gases are dissolved. Still, some of the results can be generalized for the case when the number of components exceeds two. We will obtain an asymptotic self-similar time dependence of a bubble radius and analytical expressions for the non-steady profiles of dissolved gases. We will also obtain the equation for the steady-state composition of the bubble. The steadiness of the composition is the necessary condition for the self-similar regime of bubble growth. The dependence of the steady-state composition on the solubility laws of the components will be demonstrated. For the case of strong supersaturations (which is typical for the homogeneous nucleation of a bubble) we will obtain universal expressions for the steady-state composition that are independent from the solubility laws. Since the self-similar approach to the bubble growth description will be used, we will consider relatively large bubbles. The self-similar regime of bubble growth takes place only when the influence of the surface tension could be neglected.

The diffusional growth of a two-component bubble is similar to the growth of a two-component liquid droplet in a supersaturated vapor-gas mixture \cite{Kulmala-1993, Grinin-Kuni-Lezova, Kuni-Lezova-Shchekin}. Nevertheless, there is a significant difference between these two processes. While homogeneous nucleation of droplets in supersaturated vapor occurs at relatively low supersaturations (of the order of a unit) and vapor diffusion toward a growing drop can be considered steady \cite{Grinin-Gor-Kuni-JPhysChem}, nucleation of gas bubbles in supersaturated solutions usually takes place at high values of supersaturation; therefore, as the radius of the bubble increases, the steady regime of diffusion gradually gives way to the non-steady one \cite{Kuchma-Gor-Kuni, Gor-Kuchma-Sievert}. As we will show further, the steady-state composition of the bubble depends on the degree of steadiness of diffusion fluxes of dissolved gases molecules to the growing bubble.

\section{Self-similar diffusion problems}
\label{sec:Diffusion_problems}

The state of the solution is stipulated by temperature $T$, pressure $\Pi$ and the initial densities $n_{1,0}$ and $n_{2,0}$ of the dissolved gases (hereinafter we will refer to number density of molecules using the term "density"). A gas bubble nucleates in the solution and starts growing regularly. We will consider the bubble after some time from the moment of its nucleation, i. e. when its radius $R$ has reached such a value that it satisfies strong inequality
\begin{equation}
\label{Large-R} 
R \gg 2 \sigma /\Pi.
\end{equation} 
Quantity $\sigma$ here is the surface tension of the pure liquid solvent (we consider only diluted solutions without dissolved surfactant). Eq. \eqref{Large-R} means that the influence of Laplace forces on the bubble growth dynamics is negligible; and the pressure in it is equal to the pressure of the solution $\Pi$.

The composition of the bubble is determined by concentrations of both gases in it. Under the notion of concentration we understand the mole fraction:
\begin{equation}
\label{Concentration-Def} 
c_{i} \equiv \frac{N_i}{N_1 + N_2}.
\end{equation}
Hereinafter, when index $i$ is used, we mean that the value corresponds to both cases $i=1$ and $i=2$; $N_1$ and $N_2$ are the numbers of molecules of gases $1$ and $2$ in the bubble. Evidently Eq. \eqref{Concentration-Def} is equivalent to
\begin{equation}
\label{Concentration-n} 
c_{i} \equiv \frac{n_{i,g}}{n_{1,g} + n_{2,g}}.
\end{equation}
where $n_{1,g}, n_{2,g}$ are the densities of gases 1 and 2 in the bubble,
\begin{equation}
\label{N_i} 
n_{i,g} = N_i \left( \frac{4 \pi}{3} R^3 \right)^{-1}.
\end{equation}

When strong inequality \eqref{Large-R} is fulfilled, the density profiles can be obtained by solving the following diffusion problems (see e.~g. Refs. \cite{Scriven-1959, Grinin-Kuni-Gor}
 ):
\begin{equation}
\label{Diffusion-Eq} 
\frac{\partial n_{i}(r,t)}{\partial t} = \frac{D_{i}}{r^{2}} \frac{\partial}{\partial r} \left[r^{2} \frac{\partial n_{i}(r,t)}{\partial r} \right] - \frac{R^{2}}{r^{2}} \dot{R} \frac{\partial n_{i}(r,t)}{\partial r},
\end{equation}
\begin{equation}
\label{Initial-Cond} 
\left. n_{i}(r,t) \right|_{r = \infty} = n_{i,0},
\end{equation}
\begin{equation}
\label{Boundary-Cond} 
\left. n_{i}(r,t) \right|_{r = R} = n_{i,\infty}(c_{i}).
\end{equation}
Here $r$ is the distance from the bubble center; $n_{i}(r,t)$ is the $i$-th gas density profile; $D_{i}$ is the diffusion coefficient of the dissolved $i$-th gas in the solvent (diffusion coefficients can be assumed constant for the diluted solution); $\dot{R} \equiv dR/dt$ is the rate of change of the bubble radius in time. Quantity $n_{i,\infty}(c_{i})$ is the equilibrium density of the $i$-th gas at the flat surface of the gas phase with the concentration $c_{i}$ of the $i$-th gas in it.

Even in the form \eqref{Diffusion-Eq}, \eqref{Initial-Cond}, \eqref{Boundary-Cond} the problem is non-trivial for analytical treatment. Nevertheless, if we formally assume that Eqs. \eqref{Diffusion-Eq}, \eqref{Initial-Cond}, \eqref{Boundary-Cond} are valid from $t=0$ and set the initial bubble radius value $\left. R \right|_{t=0}$ to $0$ as in Ref. \cite{Cable-Frade-Multi}
 , we can write self-similar solutions for diffusion problems, as it was previously done for the one-component case \cite{Scriven-1959, Grinin-Kuni-Gor}. It has to be noted that after the pioneering works by Zener \cite{Zener} and Frank \cite{Frank}, who used self-similar method for the description of diffusional crystal growth in the supersaturated solution, this method was exploited repeatedly by various authors for droplet growth in supersaturated vapor-gas medium \cite{Vasilyev, Grinin-Kuni-Lezova, Grinin-Gor-Kuni-JPhysChem} and for bubble growth in superheated and supersaturated solutions \cite{Scriven-1959, Lastochkin-Favelukis, Divinis-2004, Divinis-2005, Kostoglou-Karapantsios-2007, Grinin-Kuni-Gor}.

Following Refs. \cite{Vasilyev}
 and \cite{Grinin-Kuni-Gor}
  , to find a self-similar solution we introduce dimensionless variable $\rho$ as
\begin{equation}
\label{rho} 
\rho \equiv r/R.
\end{equation}
and presume that $n_i(r,t) = n_i(\rho)$. Then, Eq. for $n_i(\rho)$ can be obtained from Eq. \eqref{Diffusion-Eq} and has the form
\begin{equation}
\label{ODEs} 
\frac{d^2 n_i(\rho)}{d \rho^2} + \left[ \frac{2}{\rho} + \frac{R \dot{R}}{D_i} \left( \rho - \frac{1}{\rho^2} \right) \right] \frac{d n_i(\rho)}{d \rho} = 0.
\end{equation}
Initial and boundary conditions \eqref{Initial-Cond} and \eqref{Boundary-Cond} also can be rewritten in terms of dimensionless variable $\rho$ as follows
\begin{equation}
\label{Initial-Cond-Rho} 
\left. n_{i}(\rho) \right|_{\rho = \infty} = n_{i,0},
\end{equation}
\begin{equation}
\label{Boundary-Cond-Rho} 
\left. n_{i}(\rho) \right|_{\rho = 1} = n_{i,\infty}(c_{i}).
\end{equation}
The necessary condition for Eq. \eqref{ODEs} to have a self-similar solution is independence of fraction ${R \dot{R}}/{D_i}$ from time, i. e.
\begin{equation}
\label{RR=const} 
R \dot{R} = const.
\end{equation}
To assure whether the fulfillment of Eq. \eqref{RR=const} is possible, we need to consider the equation of material balance:
\begin{equation}
\label{Balance-i} 
\frac{dN_i}{dt} = 4 \pi R^2 D_i \left. \frac{\partial n_i(r, t)}{\partial r} \right|_{r=R}.
\end{equation}
Substituting Eq. \eqref{N_i} into Eq. \eqref{Balance-i} and transforming derivatives on $r$ to derivatives on $\rho$ [using Eq. \eqref{rho}], we obtain
\begin{equation}
\label{Balance-i-rho} 
R \dot{R} = D_i \frac{1}{n_{i,g}} \left. \frac{d n_i(\rho)}{d \rho} \right|_{\rho=1} - \frac{R^2}{3} \frac{1}{n_{i,g}} \frac{dn_{i,g}}{dt}.
\end{equation}
In order for Eq. \eqref{Balance-i-rho} to be in concordance with Eq. \eqref{RR=const}, it needs to have the second addend in its r.~h.~s. equal to $0$. It means that $n_{i,g}$ is constant in time; and this situation corresponds to the steady-state composition of the bubble. We will denote the values of these constant densities (and constant concentrations $c_i$) with superscript "$s$" -- steady-state: $n_{i,g}^s$ and $c_i^s$. Therefore, Eq. \eqref{Balance-i-rho} can be rewritten as
\begin{equation}
\label{dRdt-i-b} 
{R \dot{R}} = D_i b_i,
\end{equation}
where the dimensionless parameter $b_i$ is introduced via
\begin{equation}
\label{b-i} 
b_i \equiv \frac{1}{n_{i,g}^s} \left. \frac{d n_i(\rho)}{d \rho} \right|_{\rho=1}.
\end{equation}

Integrating Eq. \eqref{ODEs}, using Eqs. \eqref{b-i} and \eqref{Boundary-Cond-Rho}, we have 
\begin{equation}
\label{Profiles} 
n_i(\rho) = n_{i, \infty}(c_{i}^s) + n_{i, g} b_i {\rm e}^{3b_i/2} \int_{1}^{\rho }{\frac{dx}{x^{2}} {\rm e}^{-{b_i x^2/2} -{b_i/x}}}.
\end{equation} 
The value of parameter $b_i$ in Eq. \eqref{Profiles} is not known yet. To obtain this value we will use condition \eqref{Initial-Cond-Rho}, which was not exploited earlier. Substituting Eq. \eqref{Initial-Cond-Rho} in Eq. \eqref{Profiles}, we have
\begin{equation}
\label{Trans-Eq} 
a_i = b_i {\rm e}^{3b_i/2} \int_{1}^{\infty}{\frac{dx}{x^{2}} {\rm e}^{-{b_i x^2/2} -{b_i/x}}}.
\end{equation} 
where important dimensionless parameter $a_i$ is introduced using 
\begin{equation}
\label{a-i} 
a_i \equiv \frac{n_{i,0} - n_{i,\infty}(c_i^s)}{n_{i,g}}.
\end{equation}

Transcendental equation \eqref{Trans-Eq} has the following asymptotics (see Ref. \cite{Grinin-Kuni-Gor}
 ):
\begin{equation}
\label{Small-a} 
b_i \simeq a_i ~~~~~~~~~ (a_i^{1/2} \ll 1)
\end{equation} 
and
\begin{equation}
\label{Large-a} 
b_i \simeq \frac{6}{\pi} a_i^2 ~~~~~~~~~ (a_i \gtrsim 10).
\end{equation} 
Eq. \eqref{Small-a} corresponds to the situation when gas diffusion to the bubble is steady, which can take place only at low supersaturation (it is possible only when the bubble nucleates heterogeneously). Eq. \eqref{Large-a} corresponds to the strongly non-steady-state diffusion, which is typical for homogeneous nucleation of gas bubbles in supersaturated solution. For details considering steady-state conditions of gas bubble growth see Refs. \cite{Kuchma-Gor-Kuni, Gor-Kuchma-Sievert}
.

Concluding the current section we should note that describing the bubble growth by diffusion equation \eqref{Diffusion-Eq} we made two simplifications. First, we neglected the enthalpy of gas dissolution, assuming that the temperature of the bubble is equal to the temperature of the solution. The analysis of this assumption was made in Ref. \cite{Grinin-Kuni-Gor}
 for the growth of one-component gas bubble, and it was shown that this effect is negligible even at relatively high supersaturation, when $a_i \lesssim 20$.

Another simplification is related with the presupposed condition of mechanical equilibrium between the bubble and solution, in particular, we neglected the solvent viscosity. Its influence on bubble dynamics can be estimated using Rayleigh-Plesset equation (see e. g. Ref. \cite{Brennen-1995}
). For the gas pressure in the bubble $P_R$ we will have:
\begin{equation}
\label{Viscosity-1}
P_R = \Pi + \frac{2 \sigma}{R} + 4 \eta \frac{\dot{R}}{R},
\end{equation}
where $\eta$ is the dynamic viscosity of the solvent. The inertial terms in Rayleigh-Plesset equation are negligible for any reasonable bubble growth rate. In our examination we already neglected the second term in Eq. \eqref{Viscosity-1} due to strong inequality \eqref{Large-R}. To neglect the third term we need the following strong inequality to be fulfilled:
\begin{equation}
\label{Viscosity-2}
\eta \ll \frac{\Pi}{4} \frac{R}{\dot{R}}.
\end{equation}
Taking into account Eq. \eqref{dRdt-i-b}, we can rewrite this inequality as
\begin{equation}
\label{Viscosity-3}
\eta \ll \frac{\Pi}{4} \frac{R^2}{D_i b_i}.
\end{equation}

Let us estimate the value in the r. h. s. of inequality (\ref{Viscosity-3}). First of all, we need evaluation for the bubble radius $R$ from strong inequality \eqref{Large-R}. Values of surface tension both for water and for volcanic systems \cite{Navon} are $\sigma \sim 10^{-1}~N~m^{-1}$; pressure $\Pi \sim 10^5~Pa$, thus we have the minimal radius of the bubble $R \sim 2 \times 10^{-5}~m$. The value of $b_i$ is determined by the value of $a_i$ and for $a_i \sim 20$ reaches $b_i \sim 800$ (see Eq. \eqref{Large-a}). Taking $10^{-10}~m^2~s^{-1}$ as the estimation for $D_i$, we can rewrite Eq. \eqref{Viscosity-3} as 
\begin{equation}
\label{Viscosity-4}
\eta \ll 10^{3}~Pa~s.
\end{equation}
It should be noted that "common" liquids at normal conditions always satisfy this condition: for water we have $\eta \sim 10^{-3}~Pa~s$ and for glycerol $\eta \sim 1~Pa~s$ \cite{Landau-VI}. Even for volcanic systems the values of viscosity that satisfy strong inequality (\ref{Viscosity-4}) are quite typical when $SiO_2$ content is not too high (basalt, andesite and dactite melts) \cite{Sparks-1978}. However, for rhyolite melts ($\sim 70\%~SiO_2$) viscosity can reach the values of $10^7~Pa~s$ \cite{Navon}; and therefore for rhyolite melts the effect of solvent viscosity has to be taken into account.

\section{Steady-state composition of the bubble}
\label{sec:Composition}

From Eq. \eqref{dRdt-i-b} one can derive the parabolic relation between the radius and time:
\begin{equation}
\label{R(t)} 
R = \left( 2 D_i b_i t \right)^{1/2}.
\end{equation}
It has to be noted that numerical solutions of such diffusion problems showed that the assumption of parabolic relation between radius and time Eq. \eqref{R(t)} is reasonable as an asymptotic behavior for multi-component bubble for arbitrary number of gases in it \cite{Cable-Frade-Multi}.

The equality of the r.~h.~s. of Eq. \eqref{dRdt-i-b} for both cases $i = 1$ and $i = 2$ gives us
\begin{equation}
\label{D_i-b_i} 
D_1 b_1 = D_2 b_2.
\end{equation}
Eq. \eqref{D_i-b_i} is the necessary condition for the self-similarity of the two-component bubble growth. And, as we have seen, the growth of the bubble can be self-similar only when its composition ($c_1$ and $c_2$) is constant. Eq. \eqref{D_i-b_i} together with Eq. \eqref{Trans-Eq} determine the steady-state composition of a bubble. Analytical solution of these eqations for different particular cases will be presented further in this section.

It should be noted that the condition on steady-state composition of growing compound in its general form analogous to Eq. \eqref{D_i-b_i} was obtained for the case of crystal growth in the supersaturated solution \cite{Frank} and for the case of droplet growth in supersaturated vapor-gas medium \cite{Grinin-Kuni-Lezova}.

\subsection{Arbitrary supersaturations}
\label{sec:General}

We will consider two characteristic cases: when both diffusion fluxes to the bubble can be considered steady, and when both diffusion fluxes are significantly non-steady. Let us begin with the case when gas diffusion can be considered steady, and we can use Eq. \eqref{Small-a} for both $i=1$ and $i=2$. Thus, from Eq. \eqref{D_i-b_i} we have:
\begin{equation}
\label{D_i-a_i-steady} 
D_1 a_1 = D_2 a_2.
\end{equation}
Substituting definition \eqref{a-i} into Eq. \eqref{D_i-a_i-steady}, taking into account $n_{i,g} = n_{i,g}^s$ and Eq. \eqref{Concentration-n} (which gives us $c_2^s = 1 - c_1^s$), we obtain
\begin{equation}
\label{Explicit-Eq-steady} 
\frac{D_1}{D_2} \frac{1 - c_1^s}{c_1^s} = \frac{n_{2,0} - n_{2,\infty}(1 - c_1^s)}{n_{1,0} - n_{1,\infty}(c_1^s)}.
\end{equation}
Equation analogous to Eq. \eqref{Explicit-Eq-steady} was obtained by Kulmala et al. \cite{Kulmala-1993} for the steady-state growth of a two-component droplet. Eq. \eqref{Explicit-Eq-steady} gives us explicit values for the steady-state concentrations in the bubble $c_1^s$ and $1 - c_1^s$, which are unambiguously defined when the solubility laws for each gas are set. Solubility laws give us the explicit dependence of equilibrium densities $n_{1,\infty}(c_1^s)$ and $n_{2,\infty}(1 - c_1^s)$ on corresponding gas concentrations, e. g. Henry's law (which describes gases where molecules do not dissociate during dissolution)
\begin{equation}
\label{Henry} 
n_{i,\infty}(c_i) = c_i n_{i,\infty}
\end{equation}
and Sievert's law (which describes gases where molecules dissociate into two parts during dissolution)
\begin{equation}
\label{Sievert} 
n_{i,\infty}(c_i) = \sqrt{c_i} n_{i,\infty}.
\end{equation}
Here $n_{i,\infty} \equiv n_{i,\infty}(1)$ is the density of the $i$-th dissolved gas which is in the equilibrium with the bubble of the pure $i$-th gas. Substituting relations \eqref{Henry} or \eqref{Sievert} into Eq. \eqref{Explicit-Eq-steady}, it is possible to obtain the analytical solution of the latter.

\begin{figure}
\includegraphics[height=480pt,angle=270]{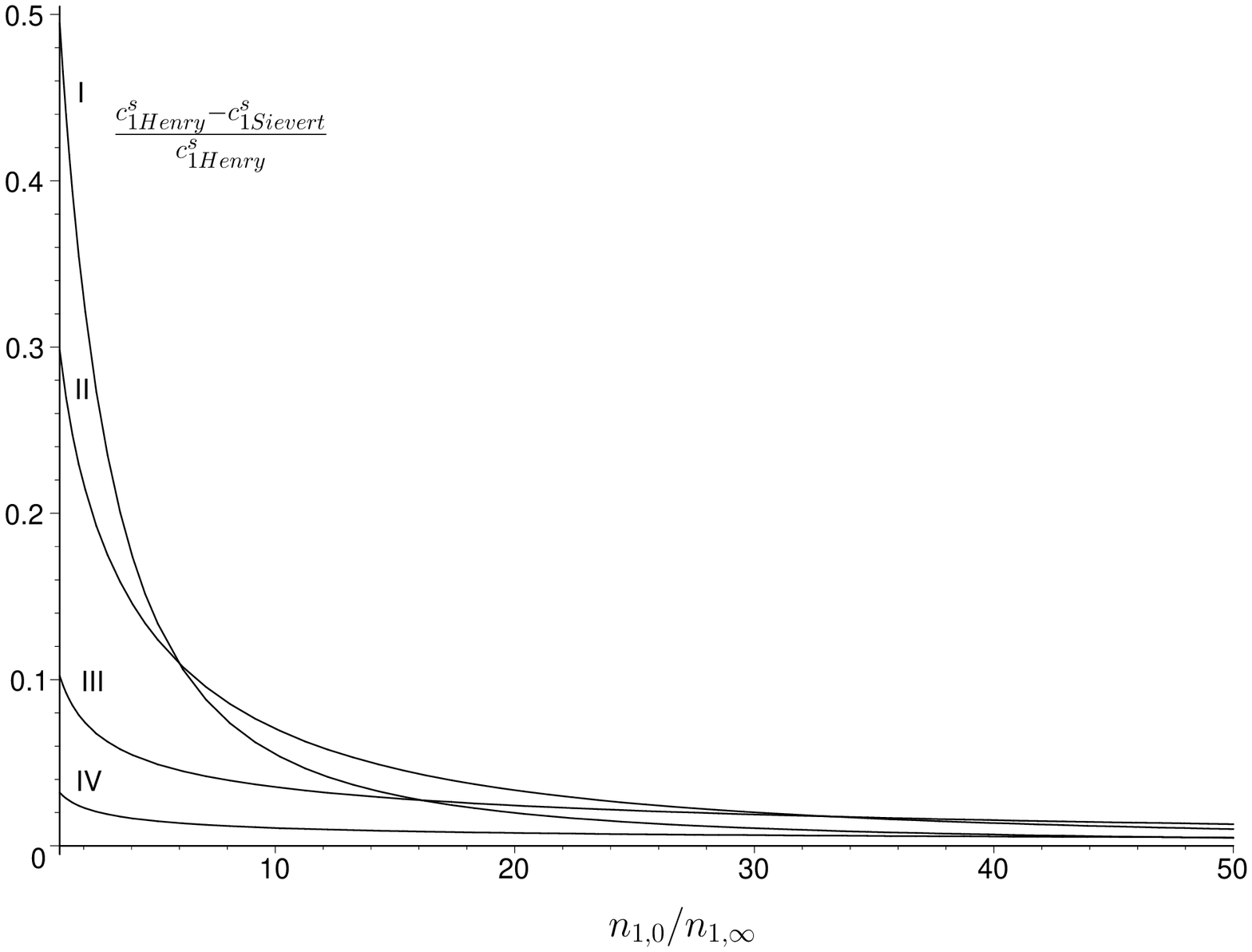} 
\caption{Relative difference between the values of steady-state concentration $c_1^s$ for Henry's law and Sievert's law as a function of $n_{1,0}/n_{1,\infty}$ for various values of $n_{2,0}/n_{2,\infty}$: curve I -- $n_{2,0}/n_{2,\infty} = 2$, curve II -- $n_{2,0}/n_{2,\infty} = 10$, curve III -- $n_{2,0}/n_{2,\infty} = 100$ and curve IV -- $n_{2,0}/n_{2,\infty} = ~1000$.}
\label{Fig:HvsS}
\end{figure}

Evidently, the steady-state composition is different for different solubility laws. To demonstrate the difference between the values of steady-state concentration $c_1^s$ for Henry's law and Sievert's law, in Fig. \ref{Fig:HvsS} we plot the value $\left( c^s_{1Henry} - c^s_{1Sievert}\right)/c^s_{1Henry}$ as a function of $n_{1,0}/n_{1,\infty}$ for different values of $n_{2,0}/n_{2,\infty}$ (for the sake of simplicity we put $D_1 = D_2$ and $n_{1,\infty} = n_{2,\infty}$). One can see that the smaller supersaturations of both components are, the higher the difference between steady-state concentrations for different solubility laws is. In the end of this section we will demonstrate that when the solution is strongly supersaturated with both components, the expression for the steady-state concentration is universal, i. e. it does not depend on solubility laws.

Let us proceed now to another case: when diffusion fluxes of both dissolved gases are significantly non-steady. Here we can use Eq. \eqref{Large-a} for both $i=1$ and $i=2$. From Eq. \eqref{D_i-b_i} we have:
\begin{equation}
\label{D_i-a_i} 
D_1 \frac{6}{\pi}a_1^2 = D_2 \frac{6}{\pi}a_2^2.
\end{equation}
Similarly to Eq. \eqref{Explicit-Eq-steady}, we have
\begin{equation}
\label{Explicit-Eq} 
\frac{D_1}{D_2} \left( \frac{1 - c_1^s}{c_1^s} \right)^2 = \left( \frac{n_{2,0} - n_{2,\infty}(1 - c_1^s)}{n_{1,0} - n_{1,\infty}(c_1^s)} \right)^2.
\end{equation}
Concentrations $c_1^s$ and $1 - c_1^s$ in the bubble can be found from Eq. \eqref{Explicit-Eq}, when the solubility laws [e. g. Eq. \eqref{Henry} or Eq. \eqref{Sievert}] are set. It should be emphasized that the applicability of Eq. \eqref{Explicit-Eq} is determined only by the condition of validity of self-similar growth of the bubble, i. e. the absence of surface tension influence. This condition does not set any limitations related with the non-steady character of the diffusion flux of dissolved gas.

\begin{figure}
\includegraphics[height=480pt,angle=270]{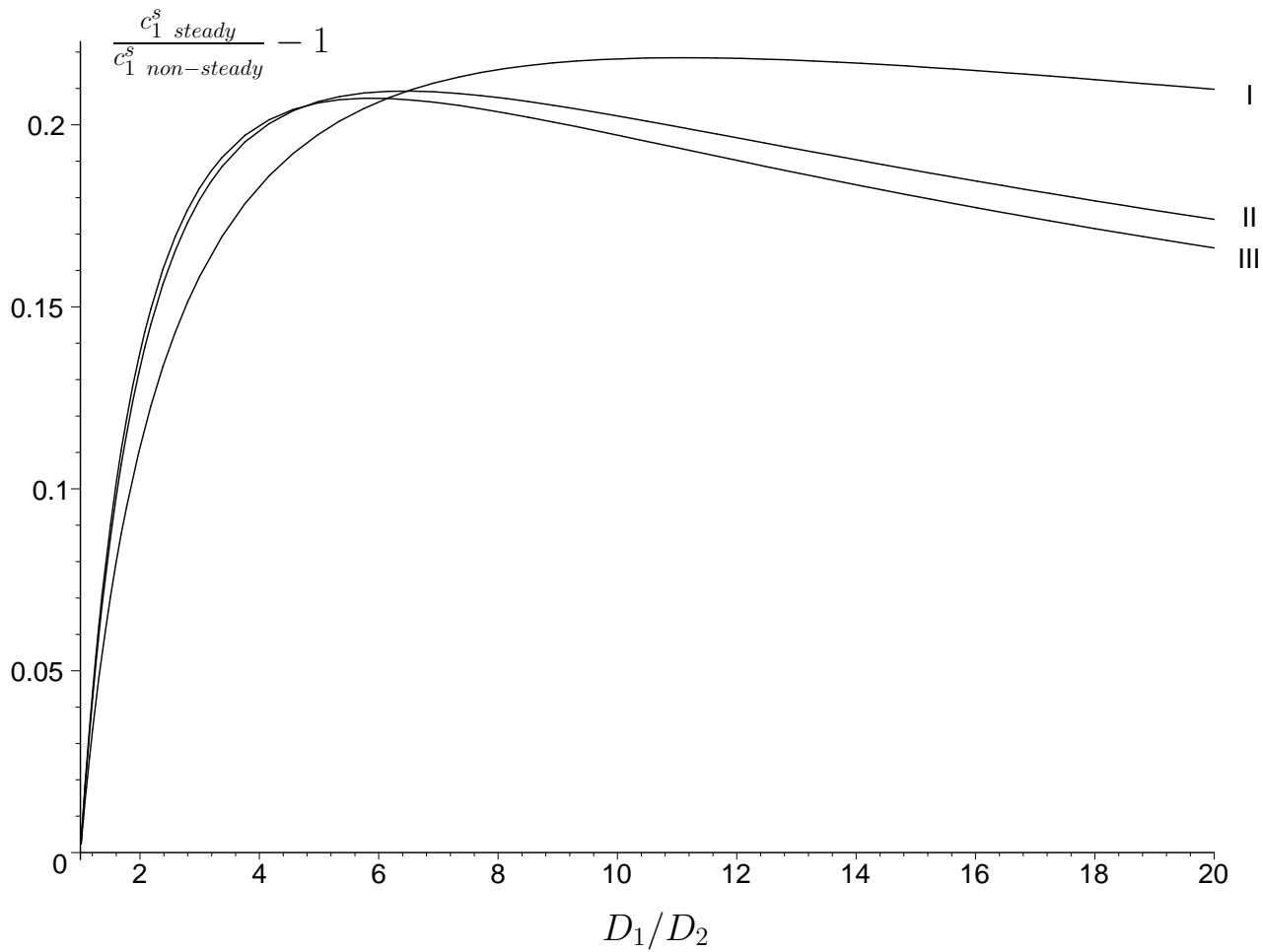} 
\caption{Relative difference between the values of steady-state concentration $c_1^s$ obtained for steady and non-steady diffusion fluxes of dissolved gases as a function of diffusion coefficients ratio $D_1/D_2$ for various values of $n_{1,0}/n_{1,\infty} = n_{2,0}/n_{2,\infty}$: curve I -- $n_{1,0}/n_{1,\infty} = n_{2,0}/n_{2,\infty} = 2$, curve II -- $n_{1,0}/n_{1,\infty} = n_{2,0}/n_{2,\infty} = 10$ and curve III -- $n_{1,0}/n_{1,\infty} = n_{2,0}/n_{2,\infty} = 100$.}
\label{Fig:SvsNS}
\end{figure}

Eq. \eqref{Explicit-Eq} was not presented in literature before; therefore, the difference between steady-state compositions of the bubble for the cases of steady and non-steady diffusion fluxes also has not been revealed previously. In Fig. \ref{Fig:SvsNS} we plot the relative difference between the values of steady-state concentration $c_1^s$ obtained for steady and non-steady diffusion fluxes of dissolved gases as a function of diffusion coefficients ratio $D_1/D_2$ for different supersaturation values.

In case when both inequalities \eqref{Small-a} and \eqref{Large-a} are violated, equation on the steady-state composition can be obtained numerically from Eqs. \eqref{D_i-b_i}, \eqref{a-i} and \eqref{Henry} or \eqref{Sievert}.

\subsection{High supersaturations}
\label{sec:High}

If we consider a solution which is strongly supersaturated with both components (such situation is typical for homogeneous nucleation of a bubble), the following strong inequality is fulfilled for both components
\begin{equation}
\label{Strong} 
n_{i,0} \gg n_{i,\infty}.
\end{equation}
Using inequality \eqref{Strong} in Eq. \eqref{Explicit-Eq-steady}, for the steady diffusion case we have
\begin{equation}
\label{Explicit-Eq-Strong-steady} 
\frac{D_1}{D_2} \frac{1 - c_1^s}{c_1^s} = \frac{n_{2,0}}{n_{1,0}},
\end{equation}
or
\begin{equation}
\label{Eq-Final-steady} 
c_1^s = \frac{1}{1 + \frac{D_2}{D_1}\frac{n_{1,0}}{n_{2,0}}}.
\end{equation}

For the non-steady diffusion case, when Eq. \eqref{Explicit-Eq} is valid, we have
\begin{equation}
\label{Explicit-Eq-Strong} 
D_1 \left( \frac{n_{1,0}}{n_{2,0}} \right)^2 = D_2 \left( \frac{n_{1,g}^s}{n_{2,g}^s} \right)^2,
\end{equation}
or, finally,
\begin{equation}
\label{Eq-Final} 
c_1^s = \frac{1}{1 + \sqrt{\frac{D_2}{D_1}}\frac{n_{1,0}}{n_{2,0}}}.
\end{equation}
It has to be emphasized that Eqs. \eqref{Eq-Final-steady} and \eqref{Eq-Final} are universal: they are valid irrespective of the solubility law, but only when the solution is strongly supersaturated with both components.

\section{Conclusions}
\label{sec:Conclusions}

While in the case of one-component bubble growth any difference in bubble dynamics for different solubility laws vanishes after $R$ exceeds $2 \sigma/\Pi$ \cite{Gor-Kuchma-Sievert}, in the two-component case the "memory" of the solubility laws is expressed in the value of the steady-state composition of the bubble even when $R \gg 2 \sigma/\Pi$.

At the same time, for strong supersaturations the steady-state composition of the bubble is universal, i. e. it does not depend on the solubility laws. For strong supersaturations the value of the steady-state concentration of each component in the bubble is defined only by initial densities of dissolved gases $n_{i,g}$ and diffusion coefficients $D_i$.

Another interesting observation is that the most complex case of bubble growth (when supersaturations are extremely high) is the simplest case for multicomponent bubble growth conditions. It is the situation when the fluxes are independent from the bubble compositions and are determined by the values of components supersaturations.

Consideration of a two-component gas bubble presented in the current paper evokes the following non-trivial question: if the bubble nucleates fluctuationally, its initial (equilibrium) composition is defined by thermodynamical parameters \cite{Baidakov-1999, Baidakov-Thermophys-2007} and does not depend on diffusion coefficients $D_1, D_2$ of the components in the solution. But at large bubble radius, as was explicitly shown above, the steady-state (but, evidently, non-equilibrium) composition of a two-component bubble does depend on these parameters. Therefore, it is important to obtain the solution to the problem of a physically correct description of a bubble evolution from the initial equilibrium composition to the steady-state one. Even for one-component bubbles the self-similar approach is not applicable when the bubble radius is small \cite{Cable-Frade, Kuchma-Gor-Kuni, Gor-Kuchma-Sievert} (when Laplace forces strongly influence bubble growth). For two-component bubbles the problem is even more complex, because the change of composition with time leads to an essential change of time dependence of the bubble radius as compared with the self-similar evolution case.

\section*{Acknowledgments}
\label{sec:Acknowledgments}

Authors are grateful to Dr. Attila Imre for stimulating our interest to the two-component gas bubbles during the brief discussion at "Nucleation Theory and Applications" workshop (Dubna, 2009).

The research has been carried out with the financial support of the Russian Analytical Program "The Development of Scientific Potential of Higher Education" (2009-2010). Project RNP.2.1.1.4430. "Structure, Thermodynamics and Kinetics of Supramolecular Systems".

\newpage 


\newpage

\newpage

\end{document}